# *Acto-myosin clusters as active units shaping living matter*

Karsten Kruse[1,*], Rémi Berthoz[2-5], Luca Barberi[1], Anne-Cécile Reymann[2-5], Daniel Riveline[2-5,*]

Affiliations:
[1] Departments of Theoretical Physics and Biochemistry, University of Geneva, 30 quai Ernest-Ansermet, 1204 Geneva, Switzerland
[2] Institut de Génétique et de Biologie Moléculaire et Cellulaire, 1 Rue Laurent Fries, 67404 Illkirch CEDEX, France
[3] Université de Strasbourg
[4] Centre National de la Recherche Scientifique
[5] Institut National de la Santé et de la Recherche Médicale

* Co-corresponding authors
Contacts: karsten.kruse@unige.ch and riveline@unistra.fr

*Abstract:* Stress generation by the actin cytoskeleton shapes cells and tissues. Despite impressive progress in live imaging and quantitative physical descriptions of cytoskeletal network dynamics, the connection between processes at molecular scales and cell-scale spatio-temporal patterns is still unclear. Here we review studies reporting acto-myosin clusters of micrometer size and with lifetimes of several minutes in a large number of organisms ranging from fission yeast to humans. Such structures have also been found in reconstituted systems *in vitro* and in theoretical analysis of cytoskeletal dynamics. We propose that tracking these clusters can serve as a simple readout for characterising living matter. Spatio-temporal patterns of clusters could serve as determinants of morphogenetic processes that play similar roles in diverse organisms.

**Introduction**

Morphogenesis is an essential part of organismal development. It is often tightly linked to cell migration, proliferation, or flows, which have been observed in different tissues during human development but also in model systems ranging from worms and flies to zebrafish and mice. These phenomena are eventually driven by the acto-myosin cortex located beneath the plasma membrane of animal cells. Given the similarities of these phenomena between many different organisms, scientists started to look for general principles guiding organismal self-organisation beyond molecular details that are specific for each system.

Acto-myosin clusters with a spatial extension of about 1µm or larger are common cortical cytoskeletal structures that have been observed in many different cell types and organisms during the last decades. Specifically, we have in mind densely connected structures of non-muscle myosin-II and actin filaments, which play a key role in determining cellular processes with important consequences for tissue shapes[1]. They are distinct from myosin mini-filaments or stacks[2]. They can form groups that extend over hundreds of micrometers and define a spatial unit[3]. In particular, contrary to myosin stacks, acto-myosin clusters lack apparent large-scale alignment of myosin-II or actin filaments. As a consequence, they should induce isotropic contraction of acto-myosin networks.

Such clusters exist in a variety of forms. First in terms of dynamics as they can be persistent or transient and they can continuously appear and disappear in a given region of the acto-myosin cortex[3,4]. In addition, their molecular composition and organisation also vary: filamentous actin in such clusters can be linear or branched and clusters containing other myosin motors, for example, myosin-I or myosin-V, also exist[5].

Despite such differences in composition and even though they are essential in some contexts, we propose that, independently of their specific composition, acto-myosin clusters form *elementary units of morphogenetic activity that pilot self-organization*. Such a point of view is not yet very common in biology, where the emphasis is typically on individual molecules. However, for understanding phenomena involving a large number of molecules, descriptions on supramolecular levels are generally appropriate[6]. In physics, so-called quasi-particles which are collective excitations in a material have been successfully introduced for describing phenomena like electric resistivity (phonons). Here, we propose that acto-myosin clusters are appropriate quasi-particles to understand morphogenetic processes.

Whereas acto-myosin clusters are commonly studied on a molecular level, in our opinion, this misses the opportunity of identifying common principles underlying morphogenetic processes. Therefore, we aim at extracting generic aspects of acto-myosin clusters without ignoring molecular details. In this spirit, we simply refer to acto-myosin clusters in this article, unless their specific composition matters. This descriptive form is also known as the mesoscopic point of view as it focuses on scales that are larger than molecular scales, but smaller than those associated with macroscopic morphogenesis. We propose that a description in terms of acto-myosin clusters together with its related properties quantitatively enable a quantitative explanation of morphogenetic processes.

The paper is organised as follows. First, we review reports on acto-myosin clusters in different biological systems, starting with fission yeast, where such clusters have originally been identified and studied. Then, we discuss their physical properties and link these to specific morphogenetic processes. We conclude by speculating how acto-myosin clusters could be informative about active mechanical stress in developing organisms and indicate how these ideas could be implemented.

**Acto-myosin clusters are a widespread structural feature of the cytoskeleton**

In this section, we show that acto-myosin clusters are indeed a common structure emerging in acto-myosin networks. First, however, we describe molecular features of acto-myosin clusters.

*Molecular properties of individual acto-myosin clusters*

At the nanometer scale, single myosin-II heads bind to actin filaments. Upon ATP hydrolysis, myosin undergoes a conformational change effectively moving it towards the actin filament's barbed end[7], Fig. 1A, B. Morphogenetic processes on cellular or tissue scales require groups of myosin-II molecules assembled in mini-filaments, which interact with assemblies of actin filaments[8], Fig. 1C. Actin filaments are additionally connected among each other by passive cross-linkers like fascin, α-actinin, filamin and others, to form crosslinked networks. Collectively, acto-myosin networks generate local mechanical stress at supra-molecular length scales from 100nm to 10µm and beyond, Fig. 1D, E.

Under the microscope, dotted patterns of fluorescently labelled non-muscle myosin-II are commonly observed, Fig. 2A, B. Each dot corresponds to a cluster as introduced above. Clusters have a minimal lifetime of 10s, but can persist for several minutes up to hours, Table I. The minimal lifetime is of the same order as that of the actin and myosin subpopulations with slowest turnover (the faster subpopulations have a lifetime of 1s or less[9]), even though in many cases internal molecular turnover is faster than the lifetime of the larger structures they compose (see Table 1). Thus clusters, as other actin networks, are dynamic steady state[10].

Acto-myosin clusters can be tracked over typically 10s or longer with standard quantitative fluorescent microscopy in essentially any morphogenetic system ranging from single cells to embryos, Fig. 2C, D and Table I. Cluster trajectories depend on the local environment, that is, the local cytoskeletal architecture and composition, Table I and [11]. Furthermore, in cytokinetic rings, cluster trajectories contain information about local mechanical stress as immobile clusters occur in contractile rings, whereas mobile clusters were proposed to act as guides for plasma membrane deposition and for the wall machinery to build the newly forming cell wall in fission yeast[12]. In both cases, the clusters contained non-muscle myosin-II.

Although we emphasise similarities between different acto-myosin clusters, it is important to keep in mind that their molecular composition can differ also for the same model system. For example, in fission yeast, cytokinetic nodes are composed of linear actin filaments interacting with myosin-II. So-called actin patches interacting with myosin-I in the context of endocytosis are composed of branched actin filaments[13]. Furthermore, myosin-V clusters are associated with linear actin bundles. Whereas clusters containing myosin-II have been observed to exhibit a wide variety of dynamic behaviours, the dynamics of actin patches and myosin-V clusters seems to be more limited, see below.

***Discovery and characterisation of acto-myosin clusters in fission yeast***

Acto-myosin clusters were first identified in the fission yeast *Schizosaccharomyces pombe*[14,15]. Called 'nodes' in this context, these myosin clusters appear in a broad region around the nucleus on the cytoplasmic side of the plasma membrane and then condense to form the cytokinetic ring[15,16], Fig. 2A. Subsequently, this ring closes, which leads to physical separation of the sister cells. Acto-myosin clusters remain discernible along the ring in the course of closure. During this process they move back and forth along the ring perimeter[12].

In addition to the late stages of cell division, acto-myosin clusters are also observed during interphase. In this phase, they locate at the tips of *S. pombe*, where they are associated with polarity cues[17]. In the same organism, actin clusters are found along actin cables, which are involved in organelle transport. However, these clusters seem to be independent of myosin and thus to be different from the clusters we focus on in this text [18].

The molecular composition of acto-myosin clusters in *S. pombe* was determined in Ref [19], where a complete list of the proteins localised in these structures along with their abundances was given. Moreover, the authors obtained the temporal sequence of arrival of the various components in the course of cluster assembly. Together with subsequent quantification[20], these measurements set the stage for grouping clusters into functional units. Furthermore, this characterisation was essential for identifying a mechanism of cluster formation.

Since their first description in fission yeast, acto-myosin clusters have been observed in the cortical actin layer beneath the plasma membrane of a large variety of animal cells and within living organisms and they have been reconstructed *in vitro* as we describe in the following.

*Acto-myosin clusters in animal cells*

In isolated cultured cells such as fibroblasts, epithelial or bone derived cells, acto-myosin clusters have been found in association with stress fibres[21], Fig. 3A, as well as with cytokinetic contractile rings[12,22,23] Fig. 3B, C, or at the ventral cortex of adherent epithelial cells, Fig. 3D. In the nematode *Caenorhabditis elegans,* myosin-II clusters appear in the actin cortex of the zygote a few minutes after fertilisation[24–26], Fig. 4A, and can be observed also at other developmental stages[27]. At the one-cell stage, quantitative analysis coupled to theory led to an understanding of their initiation and subsequent dynamics and established their role in generating and directing large scale cortical flows, which are important for specifying the embryonic anterior-posterior polarity axis and left-right asymmetry for example[24,28–30].

In the fruit fly *Drosophila melanogaster*, acto-myosin clusters form and move within a thin region below the contractile apical surface of epithelial cells during different developmental morphogenetic processes such as germband extension, gastrulation or dorsal closure[3,27,31–34], Fig. 4B. The direction of cluster movements is diverse: it is centripetal during apical contraction[35], where the clusters are implicated in cell neighbour exchanges[3], but directed towards cell-cell junctions during cell junction shrinkage[32]. Similar dynamics are reported during gastrulation in *Zebrafish*[36] Fig. 4C, adherent mammalian cells in culture[37] Fig. 3D, and in mice during neural tube closure[38] Fig. 4D. Thus, acto-myosin clusters appear frequently *in vivo*, but their dynamics as well as their roles vary.

Note that, depending on the model system, different names were given to acto-myosin clusters, for example, asters, nodes in yeast and in mammalian cells, pulses in *Drosophila*, foci in *C. elegans* or puncta in *Xenopus laevis*. Still, we emphasise similarities among these clusters as the range of their dynamic behaviour can be captured by a common framework,[40].

*In vitro reconstruction of acto-myosin clusters*

In order to identify essential components and the conditions under which acto-myosin clusters form, one can modify biochemical equilibria *in vivo* via perturbation experiments, for example, by using RNAi or mutants. However, for determining a minimum set of proteins required for spontaneous acto-myosin cluster formation, *in vitro* systems appear to provide a better approach. In these systems, one studies the formation of actomyosin clusters out of the cellular context[40–42], Fig. 5. For instance, acto-myosin clusters form spontaneously in cytoplasmic extracts[43,44] as well as in minimal systems of purified proteins containing at least actin, myosin and potentially some bundling agent[45]. The latter could either be a proper bundling protein such as anillin or fascin[46], but also Arp2/3[47], a protein that nucleates new actin filaments branching from existing actin filaments, but also myosin-II at least as ATP runs out[48].

In minimal systems, acto-myosin clusters form in a variety of geometries: in unconstrained 3D gels[49–51], on 2D lipid surfaces[40], Fig. 5A, in lipid droplets, Fig. 5B, and on actin-nucleating micro-patterned surfaces[52,53], Fig. 5C. Also, a search and capture mechanism, see below, was reconstituted *in vitro*[48], reproducing with minimal proteins the condensation of clusters during cytokinesis in fission yeast. *In vitro*, acto-myosin cluster size and dynamics depend on a range of parameters, including the order in which the various components are added, and their sizes, lifetimes, and displacement velocities span a wide range of scales including those observed in live cells or in organisms, Table I.

Clusters are thus observed within different cytoskeleton architectures and contexts, even in systems free of developmental cues. We conclude that acto-myosin cluster formation is an emerging property intrinsic to acto-myosin gels. In other words, under certain conditions, acto-myosin clusters can result from self-organisation of common cytoskeletal components and do not require neither geometrical constraints nor biochemical regulation, i.e., signalling. In this case, an initially (small) fluctuation of the acto-myosin network is sufficient to trigger their formation as positive feedback loops amplify the fluctuations and to produce clusters similar to those observed *in vivo* [26,54–57]. Let us point out, though, that up to now acto-myosin clusters reconstituted *in vitro* lack some *in vivo* properties such as pulsation and turnover: patterns once fully contracted never appear again identically at the same localisation and with the same dimensions, probably due the lack of full recycling and availability of all molecular components[40,58]. Additionally, cells have ways to regulate acto-myosin cluster formation, for example, through micro-phase separation of regulatory proteins like Mid1[59] or Cdc15[60] in fission yeast or via pulsatile patterns of signaling and active disassembly[61]. In contrast, in early *C. elegans* nascent acto-myosin clusters are actively disassembled[26].

**Physics of clusters**

Having described the widespread occurrence of acto-myosin clusters, we now turn to physical aspects of these structures. We first discuss mechanisms underlying cluster formation and then their dynamics.

*Mechanisms of cluster formation*

Optical analysis of acto-myosin cluster dynamics requires imaging at high spatial and temporal resolution. Recent advances in imaging techniques such as structured illumination microscopy allow to dissect molecular acto-myosin dynamics with unprecedented resolution. In particular, myosin motors can now be observed at the scale of individual mini-filaments moving within dynamic actin architectures[42]. Despite these advances an extensive molecular characterisation of the acto-myosin dynamics in clusters remains a challenge, also with non-optical tools.

Contrary to the case of fission yeast, in the majority of cases, we currently do not dispose of a comprehensive quantitative description of acto-myosin clusters. Even their composition has not been established in all systems. Several actin nucleation promoting factors like formins or the Arp2/3 complex are likely to contribute to the formation of acto-myosin clusters in different cellular contexts. Also, depending on the cell type, there is evidence for different bundling agents, like anillin, plastin, α-actinin or filamin, to be involved, Table I.

However, even though their molecular composition likely varies, as mentioned before, there are some important common characteristics of actomyosin clusters: their spatial extension is often around 1μm and their lifetime on the scale of minutes, Table I. The speed of moving clusters in different systems is also comparable, with a typical scale of a micrometer per minute. We report typical measurements for clusters across systems in Table I. These common features could result from similar assembly and turnover dynamics as well as from a common mechanism underlying their formation. Since acto-myosin clusters emerge in reconstitution experiments containing a relatively small number of different components, self-organisation of actin filaments, myosin motors, and cross-linkers seems to provide an appropriate frame for investigating this common mechanism. A comprehensive study of cluster formation is currently missing.

In spite of the possibility of cytoskeletal components to self-organise into clusters, the formation of acto-myosin clusters *in vivo* could be intimately linked to signalling pathways, notably the Rho GTPase pathway[34,63–65]. However, instead of regulating cluster formation in detail, based on the spontaneous emergence of clusters *in vitro*, we speculate that signalling could generate appropriate local fluctuations of the actin and myosin densities or myosin phosphorylation levels, which would then trigger cluster self-organization. In this article, we do not consider these upstream pathways and rather refer to other reviews about the connection between signalling and cytoskeletal dynamics[66,67].

*Theoretical analysis of cluster dynamics*

Theoretical physics offers a powerful approach to test possible mechanisms underlying the formation of acto-myosin clusters, Fig. 6. Detailed simulations of all molecules involved are currently typically not possible on the time scales relevant for cluster formation. A good understanding of mechanisms can often be achieved by using effective descriptions that capture molecular details by a few phenomenological terms. Effective descriptions can be introduced on different levels. One approach is through simulations that still consider individual filaments, but with an effective dynamic that does not explicitly refer to the detailed action of motors or nucleation promoting factors. This approach has been used to establish a search and capture process for condensation of clusters in *S. pombe*[68], Fig. 6A. This analysis shows that, indeed, this autonomous mechanism can promote the condensation of clusters into a ring, see also[55], Fig. 7.

Another powerful approach is called generalized hydrodynamics. In this framework, one aims not at describing individual molecules, but directly their collective behaviour. This is similar to usual hydrodynamics for describing water, where the water density and collective flows are considered rather than individual water molecules. The corresponding equations are derived based on the symmetries of the system (see Box Theory). Applied to the cytoskeleton, such a description can yield fundamental general insights into cluster formation and dynamics[69], Fig. 8. Furthermore, by considering the same read-outs as in experiments, one can also aim at a quantitative description of acto-myosin clusters, see for example Fig. 6B and Ref. [70]. The latter requires to measure or estimate experimental quantities whose values are injected as parameters into the theory, for example, those given in Table I.

Beyond reproducing observed structures in specific situations, further insights can be gained by exploring theoretical outcomes when these parameter values are varied. This leads to so-called phase diagrams, Fig. 7 and Fig. 8, which represent the various regions in parameter space, where clusters exist, where clusters are mobile or immobile[12, ,41,56,72,73]. Experimental versions of phase diagrams can be achieved for example by genetic modifications, such as the expression of a constitutively active protein, or by exposing the system to drugs in various concentrations, which correspond to modifications of parameter values. By focusing on qualitative state changes, phase diagrams offer a robust possibility for testing a theory, whereas accurate measurements of parameter values are often difficult or sometimes impossible. However, the link between experimental modifications and changes of parameter values in a theory can be difficult to establish. Also, one should keep in mind that distinct molecular mechanisms can map onto the same effective description, see Fig. 6C. For example, different molecular processes can lead to stress generation in an acto-myosin network[11] that are all captured by the same formal expression. Another example is the formation of clusters which can be the consequence of active motor sorting or density-dependent feedback on contractility among others[61,73,74].

In the simplest version, a physical description treats the cytoskeleton as an isotropic active fluid. Here, 'isotropic' refers to neglecting a possible alignment of actin filaments, whereas 'active' refers to the generation of mechanical stress, for example, through the action of molecular motors. In this case, clusters can form if the motor-generated stress exceeds a threshold value[75]. These clusters merge and eventually all the material is concentrated in one structure. In these descriptions, clusters emerge because the active stress increases with the actomyosin density, such that actomyosin continues to accumulate in regions with higher density [71,76].

Several mechanisms can counteract cluster growth. This is, for example, the case when turnover of actin filaments is accounted for[52], or in presence of a regulator, such as, for example, myosin light chain kinase, that is advected with flows of the acto-myosin network[72,78,79]. Depending on the processes that are accounted for by the theory, different dynamic states can emerge spontaneously. For example, clusters can be transient, oscillatory, or migratory[12], Fig. 8B. Descriptions that include filament alignment can yield clusters with a rich internal structure[23,39].

Finally, theory can be used to investigate possible roles of acto-myosin clusters in cellular processes. Functions can be associated to these dynamics. For example transient clusters have been proposed to play a role in transmitting information quickly in cells[80]. By focussing proteins in clusters, active forms can efficiently activate passive proteins. Dispersion of these active proteins and the subsequent reformation of clusters containing again active and inactive proteins could spread activity faster in the cell cortex than by diffusion. Experimental evidence for this mechanism has been reported, for example in Refs. [69] and [81], and theoretical work reproduces self-assembly of these clusters[69].

Node condensation in fission yeast or pulsatile contractions in cells and in *Drosophila* are recapitulated from these mesoscopic approaches. The changes in scales are not a challenge since these theories can lead to dimensions which can be tuned according to their associated physics.

It would be attractive to connect molecular organisation to mesoscopic scales up to macroscopic morphogenesis. We want to point out however that different molecular mechanisms can lead to the same macroscopic phenomenon. Notably different types of molecular motors in conjunction with actin and various actin binding proteins can all lead to tissues flows and shape changes[11]. However as pointed out by Anderson in his article 'More is different'[6], for describing a macroscopic phenomenon there is an appropriate scale – and it is not molecular. We propose that acto-myosin clusters are at the relevant scale to understand morphogenetic events like flows and shape changes.

**Shape changes driven by networks of acto-myosin clusters**

Based on the material reported so far, we suggest that actomyosin clusters provide fundamental elementary units that act as precursors for various cell shape and tissue changes. Furthermore, we surmise that their dynamics provides the nexus between molecular processes and morphogenetic processes. We will illustrate this idea with two examples, namely T1 transitions in *Drosophila* and cytokinetic ring closure in fission yeast and mammalian cells. Other notable examples are the roles of clusters during bleb retraction[82] and their potential contribution to stress generation during amoeboid motility[83].

*Flows during development*

In *Drosophila*, it was reported that acto-myosin clusters could oscillate between the cytoplasm and the ventral-dorsal junctions of the *amnioserosa* cells[32]. This back-and-forth motion of the order of few micrometres per minute could be measured in cells of the same epithelial tissues. Over the timescale of about 30 minutes, clusters accumulated within the dorsal-ventral cell-cell junction. Subsequently these clusters self-organised within the junctions and this led to a decrease in their size realising the first step of a T1 transition[32], Fig. 4B.

This example shows that cluster dynamics can be tightly correlated to the local changes of shapes at the cell-cell interfaces. Recruitment of acto-myosin clusters leads to the reduction in size of the junction. Also, their motion could set a timescale by periodically oscillating and mechanically pulling the cell-cell interface. The timescales of oscillations were involved in determining the length scales at which neighbour exchanges occurred. Interestingly, these back-and-forth dynamics associated with coupling to the cell-cell junctions were reported in other systems like in *C. elegans* with similar consequences for morphogenesis[27].

This *in vivo* example illustrates how the tracking of acto-myosin clusters can serve as a readout to quantitatively predict morphogenesis from simple movies of developing embryos. Each cluster can be followed with live microscopy and the corelation between clusters motion could serve as determinants for the shape transformations.

*Cytokinesis*

The basic machinery for cytokinesis has been identified in fission yeast and was successfully transposed to other systems such as in sea urchin[1,22] and *C. elegans*[29]. In the following, we illustrate spatio-temporal patterns of clusters during cytokinesis in different species. In these species, cytokinesis exhibits common features, in particular, the presence of clusters of similar spatial extensions. However, their dynamics can be distinct in different systems.

By placing cells vertically in cavities, cytokinetic acto-myosin rings in fission yeast and mammalian cells were visualised in a single plane of focus[12]. This method allowed to capture the dynamics of acto-myosin clusters. In fission yeast, acto-myosin clusters moved back and forth along the ring, whereas in mammalian cells they remained immobile along the ring during closure. This finding was surprising as the molecular composition of the acto-myosin clusters was thought to be the same in both cases. Using physical theory, it was found that moving and static acto-myosin clusters can be generated by the same mechanism[12,72]. The different migratory behaviour yielded differences in the mechanical stress they generated along the ring: mobile clusters corresponded to low stress, immobile clusters to large stress. In agreement with this result, experimental modification of actin dynamics and myosin activity allowed to change between these two moving and still actomyosin clusters[12]. These results about the translation between dynamics of clusters and stress generation should not be specific to the cytokinetic rings but valid for many systems.

**Conclusions**

We have reviewed experimental and theoretical studies together showing that self-organised acto-myosin clusters in a wide range of species behave locally and globally according to common rules. We propose that each system determines its points of control in a specific manner to lead the right transformation at the right time such as pulsations for T1 transition in *Drosophila* or still clusters within acto-myosin rings during cytokinetic contraction in fission yeast for example. These classes of transformations would gain in being also classified in terms of biological functions. Flowing or pulsating, or fluidising tissues could lead to categories which could emerge from clusters collective effects. Tuning cellular parameters could be the coupling across scales through adhesions secured by the local recruitment of active or passive cross-linkers. If the physical phases in phase diagrams could be conserved, the molecular mechanisms involved in the generation of clusters could be distinct depending on cellular contexts and tissues. In all cases, the systems need to be resistant to fluctuations in protein concentrations to secure morphogenesis. It will be interesting to test these ideas in embryos while outlining the mechanisms securing robust morphogenesis with outstanding precisions over time and space.

Apart from their biological significance, we speculate that acto-myosin clusters can also be used as read-outs for physical parameters. In particular, the analysis of their trajectories might be used to infer stress patterns in cells or tissues. As such, we propose that acto-myosin clusters might act as appropriate quasi-particles on which general principles underlying morphogenesis can be built. It will be interesting to test these ideas in embryos while outlining the mechanisms securing robust morphogenesis with outstanding precisions over time and space.

*Acknowledgements: We thank Julien Vermot, Pierre-François Lenne and Thomas Lecuit for insightful discussions. RB is supported by IMCBio. This work of the Interdisciplinary Thematic Institute IMCBio, as part of the ITI 2021-2028 program of the University of Strasbourg, CNRS and Inserm, was supported by IdEx Unistra (ANR-10-IDEX-0002), and by SFRI-STRAT'US project (ANR 20-SFRI-0012) and EUR IMCBio (ANR-17-EURE-0023) under the framework of the French Investments for the Future Program. KK, RB, LB and DR are supported by SNSF Sinergia grant CRSII5_183550.*

### *Box 1: Force measurements*

*Many reviews report methods to measure forces* in vitro *and* in vivo[84,85]. *Briefly, several approaches are reported with major results giving forces with relevant orders of magnitude, typically nanonewtons at the level of micrometre of active matter.*
- *Fluorescence Resonance Energy Transfer (FRET) sensors: fluorescent local signals are correlated to local forces after calibration.*
- *Laser ablation: cellular structures are locally destroyed with a focused laser and the opening speed is measured. The approach is non-invasive but assumes values for the local cell friction.*
- *Micrometer size bead and droplets: beads or droplets are placed on or within cells, and they are trapped and tracked with optical tweezers or magnetic tweezers. Both approaches have led to the determination of rheological properties of cells with imposed deformation or force.*
- *Atomic Force Microscope (AFM): this method allows to measure local forces on the outer layer of cells and tissues at the scale of micrometer.*
- *If cells and tissues can be placed on deformable substrates, the local deflection can inform about local and global force generations correlated with cellular events.*
- *Direct optical trapping: dense matter can be deformed locally and its recoil can be analysed. These measurements can give estimates for local forces.*
- *Cells and tissues traction experiments: cells and tissues are extended mechanically either with constant elongation or constant forces, and the associated curves inform about the global response of cells and tissues.*

- *Flows: with light, liquid flows were induced to study the local and global reorganisation of tissues which can give force distributions within cells and tissues.*

### *Box 2: Theory of active gels for morphogenesis*

Active gel theory[86] provides a framework for studying mechanisms underlying the formation of acto-myosin clusters. The state of the acto-myosin network is captured by its density $\rho$ and its local orientational order $\boldsymbol{p}$, that is, the degree of local alignment of the actin filaments. Its dynamics is expressed through the cytoskeletal flow velocity $\boldsymbol{v}$ and changes in the orientational order.

Cytoskeletal flow is generated by gradients in the mechanical stress $\sigma$. This stress results from pressure in the network, viscous dissipation and from processes that convert chemical energy into mechanical work commonly referred to as active processes. Examples of the latter are the action of myosin motors and the polymerisation of actin filaments. In phenomenological descriptions of active gels, however, the stress generating processes are not specified, but captured through phenomenological terms. Associated with these are phenomenological parameters like the viscosity $\nu$ or the activity parameter $\zeta$ that quantifies how much stress is generated by all active processes given a difference $\Delta\mu$ in the chemical potentials of ATP and its hydrolysis products. The values of the phenomenological parameters typically depend on the molecular composition of the cytoskeleton, whereas their existence is independent of the specific proteins involved[87].

An important task is to relate these phenomenological quantities to experimental read-outs. To this end, the dynamic fields $\boldsymbol{p}$, $\boldsymbol{v}$, $\sigma$, and $\Delta\mu$ have to be measured, from which the values of the phenomenological parameters $\zeta$ and $\nu$ (and others) can be inferred[88]. The filament alignment $\boldsymbol{p}$ can be obtained from fluorescence images of actin filaments or microtubules by the structure tensor method that is, for example, available as an ImageJ plugin. In an isotropic network, we have $\boldsymbol{p} = 0$. The flow $\boldsymbol{v}$ and hence its gradients $\nabla \boldsymbol{v}$ is accessible through particle image velocimetry measurements. Typical values for cytoskeletal flow velocities are in the range of micrometre per minute.

The measurements of both, alignment and velocity, are non-invasive assuming the fluorescence labelling does not alter the system dynamics. In contrast, measuring mechanical stress requires either to embed probes (stress sensors) into the cytoskeleton or to locally deform the filament network. In the first case, deformations of the probes, for example, oil droplets[89], together with the knowledge of their mechanical properties directly yield the stress. In the second case, the deformation of the cytoskeletal network itself serves as a proxy for the stress. For example, deformations of the actin network have been recorded after local cuts of the cortex by laser ablation[90] (Box on Experimental approaches). Typical values of the total mechanical stress are in the range of kilopascals.

The remaining quantity $\Delta\mu$ could in principle be assessed by standard techniques from chemical physics, which, however, would need to be adapted to cellular systems. For the time being, such a measurement has not been performed. Instead, there are estimates of the active stress $\zeta\Delta\mu$, which again yielded values in the range of tens of kPa[88,91,92]. Using $\Delta\mu = 30 \ kJ/mol$, we estimate $\zeta$ to be of the order of 1mM. Interestingly clusters in fission yeast were reported to have this concentration of myosin II[19]. This concentration corresponds to the densely packed configuration of molecular motors of typical size 1nm.

## *Figure captions*

### *Figure 1 - From molecules to clusters.*
*Acto-myosin structures and interactions at different length scales. (A) Illustration of a single myosin-II motor interacting with a polar actin filament. (B) Myosin decoration on single actin filament [7]. (C) Illustration of two filaments cross-linked by a myosin-II minifilament. (D, E) Illustration (D) and electron micrograph (E) of acto-myosin clusters[8].*

### *Figure 2 – Clusters and their dynamics.*
*(A) Illustration of fluorescent images of clusters. (B) Fluorescence microscopy image of myosin clusters in S. pombe[68]. (C) Illustration of potential cluster dynamics observed with fluorescence microscopy. (D) Representation of the dynamics in illustrated in (C) in form of kymographs. Blue lines: individual stochastic tracks; dotted lines: average behaviour. Colour code on schemes: blue: actin, red: myosin.*

### *Figure 3 – Actomyosin clusters across* ex vivo *systems.*
*Fluorescence microscopy images and illustrations of actomyosin clusters in various organisms. (A) Actin stress fibres in HeLa cells without (left) and with (right) 200 nM Latrunculin A treatment[21] (F-actin is labeled). (B) Myosin organised in clusters around the cytokinetic ring in fission yeast seen from a plane perpendicular to the ring before (left) and during (center) cytokinesis, and seen from a plane parallel to the ring during cytokinesis (right)[19]. (C) Myosin clusters in mitotic HeLa cells during cytokinesis, seen from a plane parallel to the ring[12].*

### *Figure 4 – Acto-myosin clusters across* in vivo *systems*
*(A) Actin (red) and myosin (green) in one-cell C. elegans embryos[4]. (B) Myosin (green) and membrane (red) during* Drosophila *germband elongation[3]. (C) Myosin (left) and actin (right) during zebrafish gastrulation[36] (EVL, enveloping cell layer surface epithelium, YSL, yolk syncytial layer). (D) Myosin (green) and cadherin (purple) during mouse neural tube closure for wild-type (Grlh2$^{+/+}$) and Grlh2 mutants (Grlh2$^{-/-}$) (surface ectoderm (se)/neuroepithelium (ne) boundary). Mutants fail closure and lack myosin clusters[38]. Illustrations highlight the clusters in the corresponding fluorescence images. Colour code as in Fig. 2.*

### *Figure 5 – Acto-myosin clusters across* in vitro *systems.*
*(A) Acto-myosin systems reconstituted* in vitro *([93](top); [40](bottom)). Green: myosin, red: actin, blue: capping protein. (B) Acto-myosin system reconstituted in liquid droplets[94]. Purple: actin. (C) Closing acto-myosin formed by micropatterning[53]. Color-coded kymograph showing constriction over time. Illustrations highlight the clusters in the corresponding fluorescence images. Colour code as in Fig. 2.*

### *Figure 6 - Generic molecular rules for clusters.*
*(A-C) Generic rules of interactions as basis for theoretical approaches, search and capture for cytokinesis[68] (A), bipolar structures with actin of opposite polarities projecting from nucleators[12] (B), aster formation[55] (C).*

### *Figure 7 – An example of acto-myosin clusters in a molecular theory of cytoskeletal dynamics.*

(A) Snapshots of a stochastic simulations of actin filaments and molecular motors with turnover for different turnover rate $p_2$ and viscosities $\eta$[55]. (B) Phase diagram as a function of filament turnover and motor-filament work.

***Figure 8 – Clusters in effective theories of cytoskeletal dynamics.***
*(A) Hydrodynamic description of acto-myosin as a two-dimensional, polar active fluid[69]. Left, specific solutions for different parameter values. Right, phase diagram. (B) Kinematic description of acto-myosin dynamics in the cytokinetic ring, where filaments can form bipolar structures[12]. Left, kymographs of specific solutions, featuring static and moving clusters. Right, phase diagram.*

***Table I.***
*Typical spatial and temporal scales for clusters together with their molecular compositions. The data is grouped for (i) single cells, (ii) embryos, (iii) in vitro reconstitution.*

*Table 1 – A non-exhaustive list of clusters and their characteristics. Molecular turnover measured by methods such as FRAP do not correspond to the lifetime of clusters which suggests that clusters are dynamic steady states (note that myosin nomenclature should be taken with due care between organisms, see for example Ref. [5]).*

| Organism / Experimental conditions (Clusters name) | Diameter | Spacing | Velocity | Lifetime (assembly-disassembly) | Molecular turnover | Composition | Actin filament orientation | References |
|---|---|---|---|---|---|---|---|---|
| Yeast *Schizosaccharomyces pombe* (Nodes) | < 1 μm | < 1 μm | 0.1 to 0.5 μm/s | long lived, stable for +10 min | actin ≈ 1 min | anillin Mid1p, Myo2, IQGAP, Rng2p, F-BAR, Cdc15p, formin Cdc12p | barbed filaments inwards (formin in nodes) | Pollard and Wu, 2010[14] |
| Yeast *Saccharomyces cerevisae* (Patches) | 1 μm | 1 μm | 0.1 to 0.5 μm/s | 1 min | actin 10 to 20 s | Arp2/3 at apex, Abp1, Myo3, Myo5 | uniform polarity, barbed ends facing cortex | Moseley et al, 2006[5] |
| Yeast *Schizosaccharomyces japonicus* (Cell ghosts) | 0.46 to 1.15 μm | 1 to 4 μm | static | 1 h | | Rlc1, Cgc15, Myo2, Myp2, Rng2 | pointed ends outwards | Chew et al, 2017[95] Thiyagarajan et al, 2022[55] |
| MEF, HeLa cell cultures / LatA treatment (Nodes) | 1 μm | 1 to 2 μm | < 1 μm/min | 300 s | | formin DAAM1, crosslinkers/filamin FlnA, myosin | barbed ends inwards (formin in nodes) | Luo et al, 2013[21] |
| MCF-7, MEF, U2OS (Pulses) | 3 μm | 1 to 2 μm | < 1 μm/min | 60 s | | NM2A, NM2B | | Baird et al, 2017[37] |
| *Drosophila* / Germband (Clusters) | 1 to 5 μm | 2 to 5 μm | 6 μm/min | 1 min | | | | Rauzi et al, 2010[62] |
| *Caenorhabditis elegans* / Zygote | 3 to 4 μm | 5 μm | 7 μm/min | 30 s | actin 5 to 10 s myosin 6 to 17 s | formin CYK-1 or Arp2/3 depending on the stage, NMY-2, anillin (center), plastin (aster) | barbed ends moving outwards | Nishikawa et al, 2017[97]; Robin et al, 2014[96]; Yan et al., 2022[26]; Michaux et al, 2018[29]; Costache et al, 2022[28] |
| *Zebrafish* / Gastrulation | 1 μm | 2 μm | 1 μm/min | > 1 min | | | | Behrndt et al, 2012[36] |
| *Xenopus* / Egg extract on oil droplet | 5 μm | 10 μm | 10 μm/min | 1 h | | | | Tan et al, 2018[44] |
| Proteins, Lipid bilayer | 2 to 5 μm | 5 to 10 μm | 0.5 to 1 μm/min | 9 min | | actin, myosin, Capping Protein +actin binding domain of ezrin (actin membrane linker/tether) | barbed ends inwards (myosin at the center) | Köster et al, 2016[40] |
| Proteins, bulk | 30 to 1000 μm | 30 to 1000 μm | static | stable once assembled | ≈ 5 s | actin, myosin, crosslinker required | | Alvarado et al, 2013[49] |
| Proteins, micropatterning | 2 to 5 μm (imposed) | 2 to 10 μm (imposed) | max 0.8-2.6 μm/min | 15 to 30 min | | Arp2/3, α-actinin (optional), myosin-V dimers (other also possible) | mixed polarity at the center, pointed ends outwards | Reymann et al, 2012[53]; Ennomani et al, 2016[52] |

Figure 1

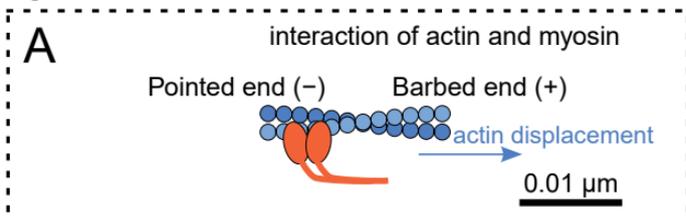

A — interaction of actin and myosin
Pointed end (−)   Barbed end (+)
actin displacement
0.01 µm

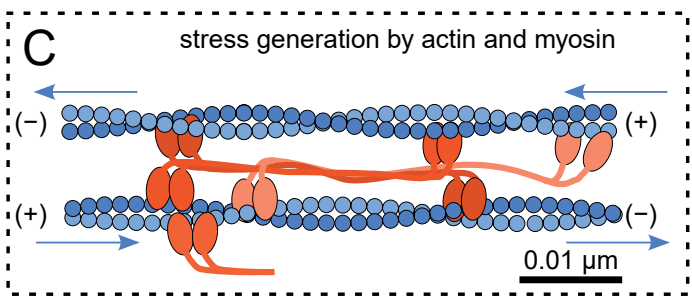

C — stress generation by actin and myosin
(−) (+)
(+) (−)
0.01 µm

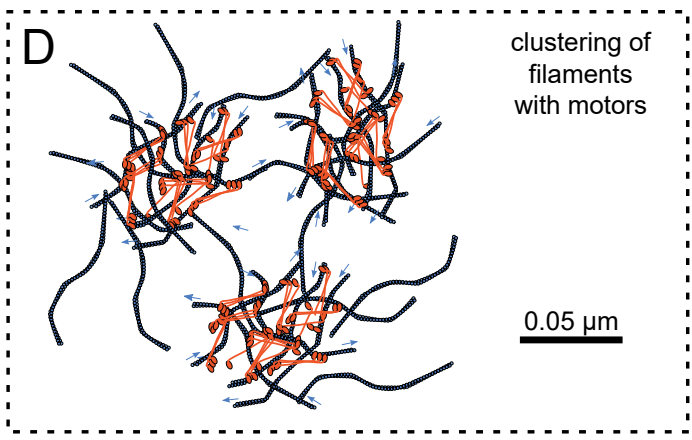

D — clustering of filaments with motors
0.05 µm

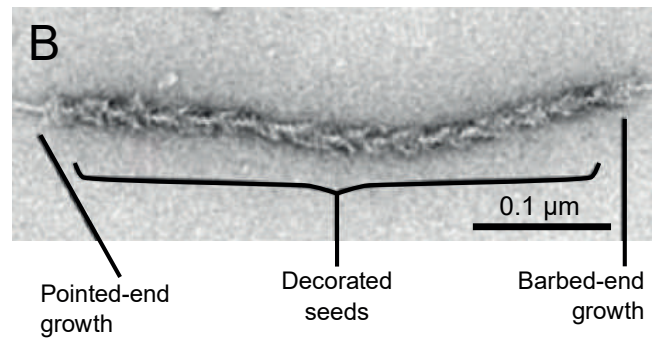

B
Pointed-end growth   Decorated seeds   Barbed-end growth
0.1 µm

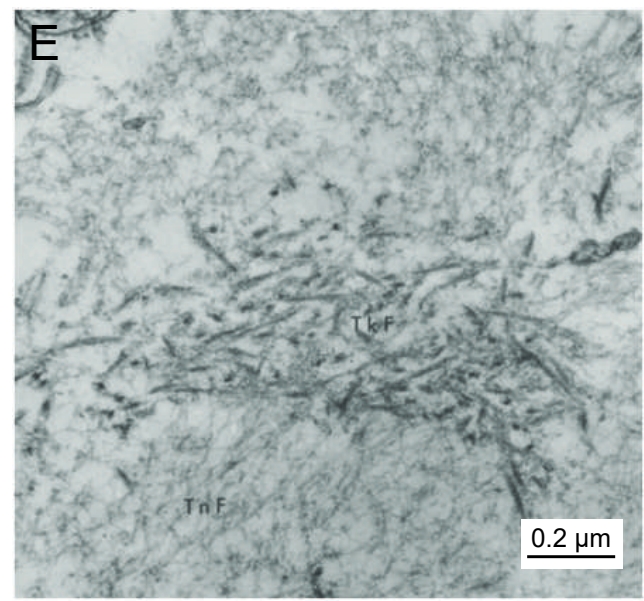

E
TkF
TnF
0.2 µm

Figure 2

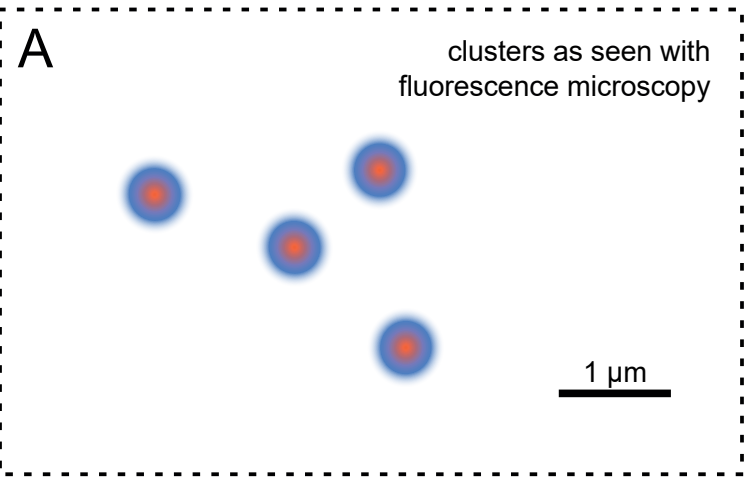
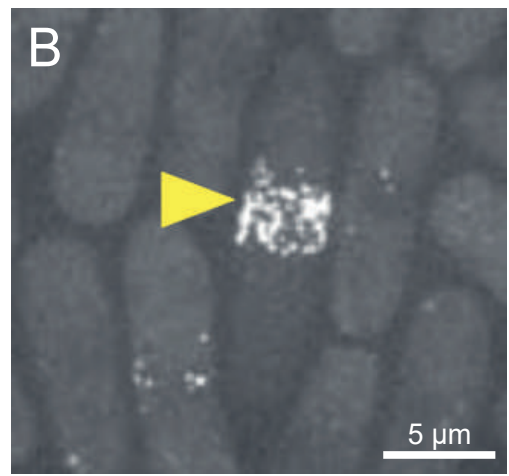
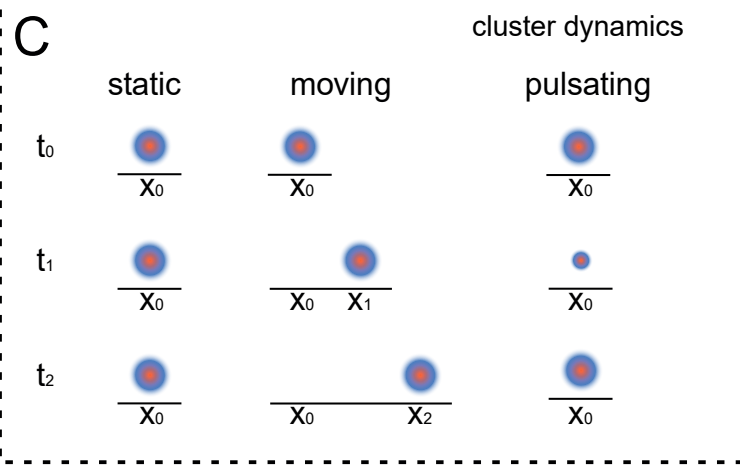
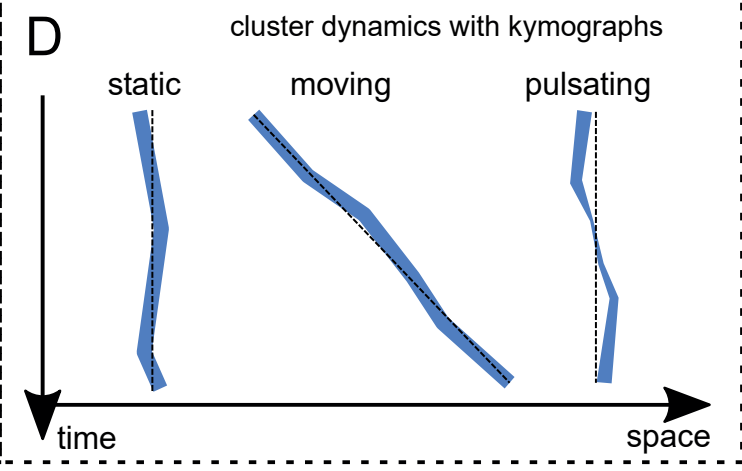

Figure 3

A HeLa cells - actin nodes in stress fibres

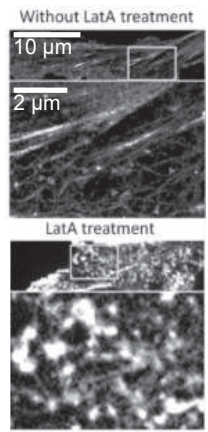

B Fission yeast - nodes forming cytokinetic ring

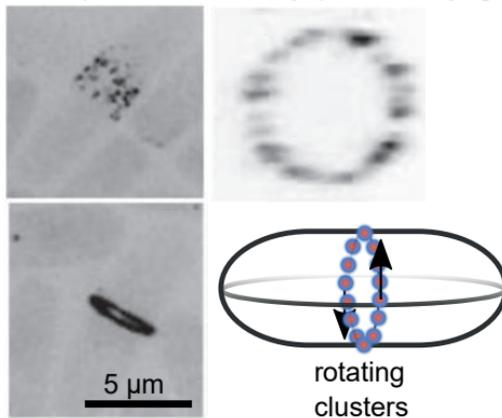

rotating clusters

C HeLa cells - clusters within cytokinetic ring

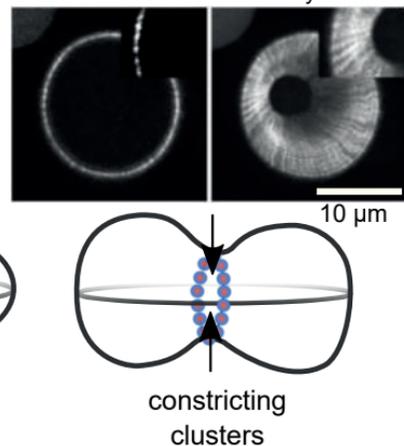

constricting clusters

D Adherent cells in culture - pulses at the ventral cortex

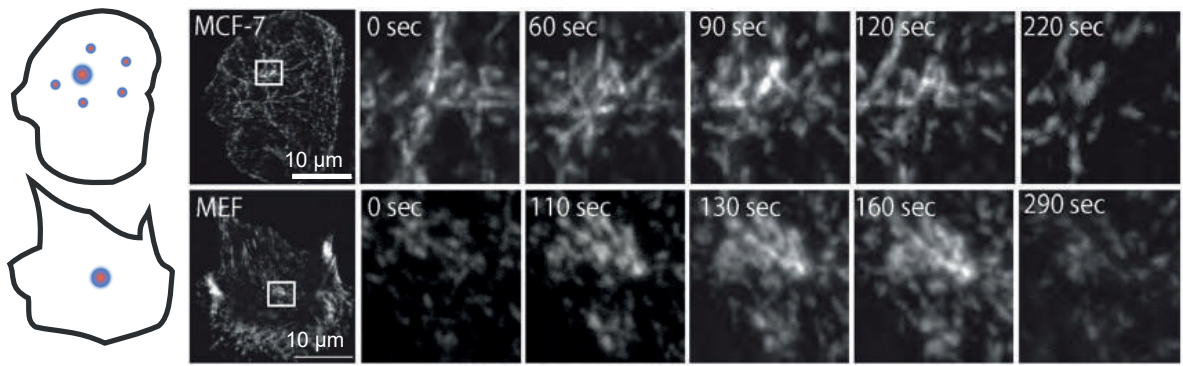

## Figure 4

### A *C. elegans* zygote - cortical pulses and contractile ring

### B *Drosophila* - pulsating clusters in germband extension

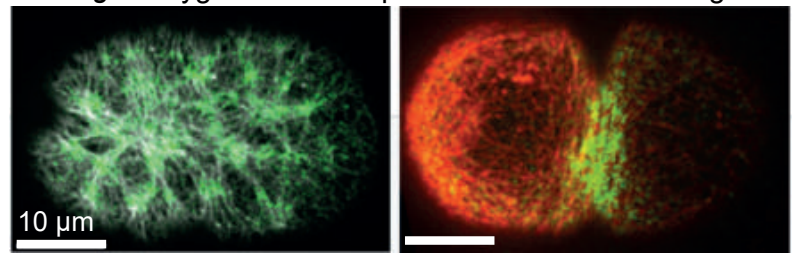
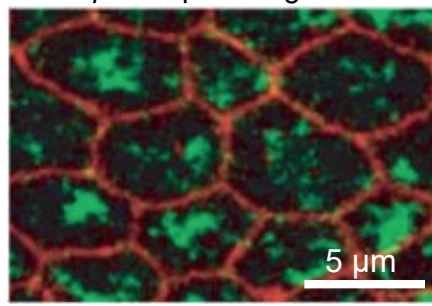
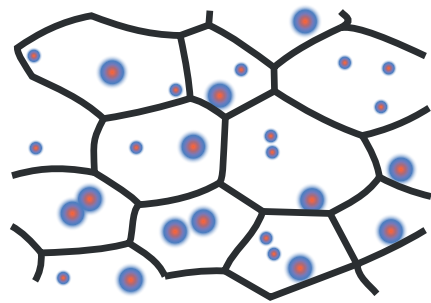

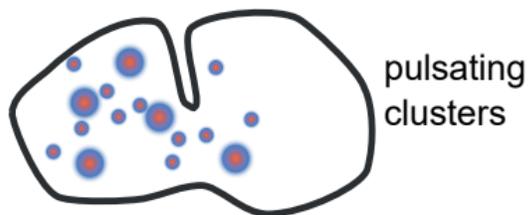

pulsating clusters

### C *Zebrafish* gastrulation

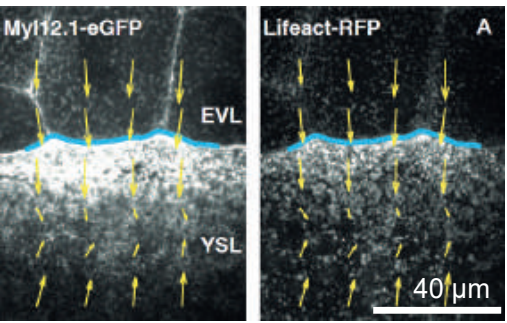

moving clusters

### D Mouse - clusters in neural tube closure

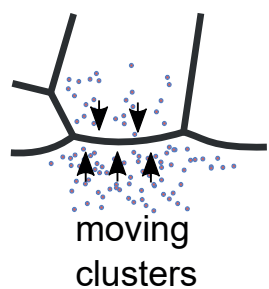
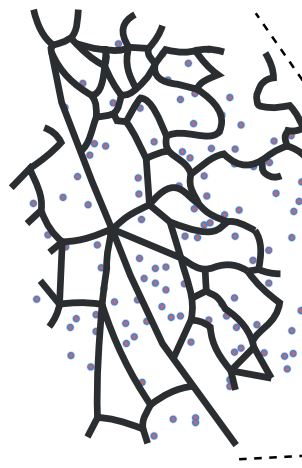
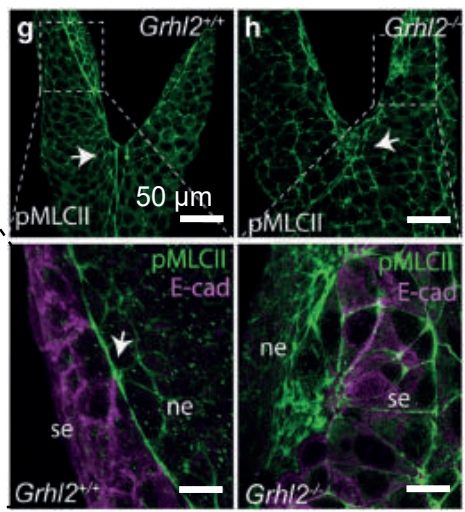

Figure 5

**A** *In vitro* reconstitution on lipid bilayer - myosin-induced formation of actin asters

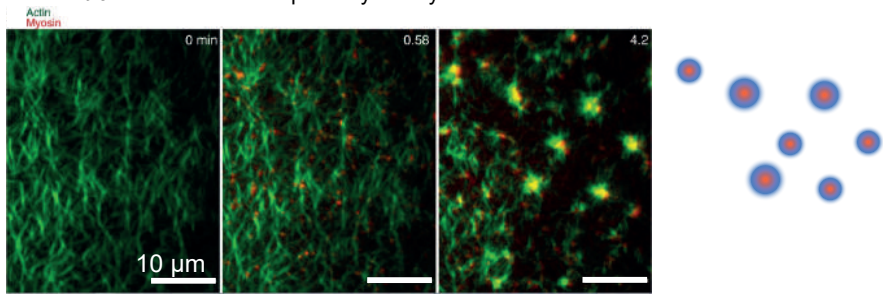

**B** *In vitro* reconstitution on lipid droplets - myosin-induced formation of actin puncta

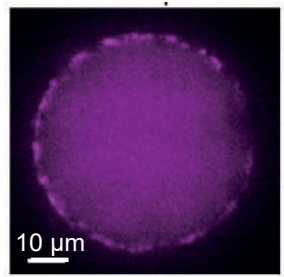

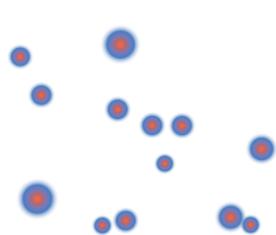

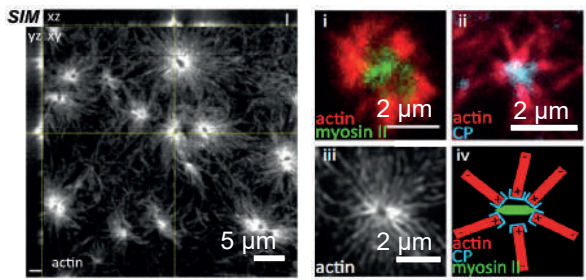

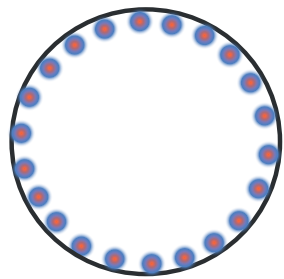

**C** *In vitro* micropatterning of actin and myosin ring - closes over time

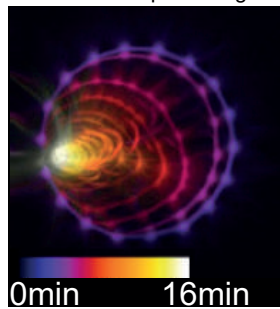

*kymograph, the ring closes over time*

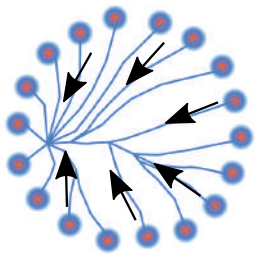

Figure 6

A Search, capture, pull and release model

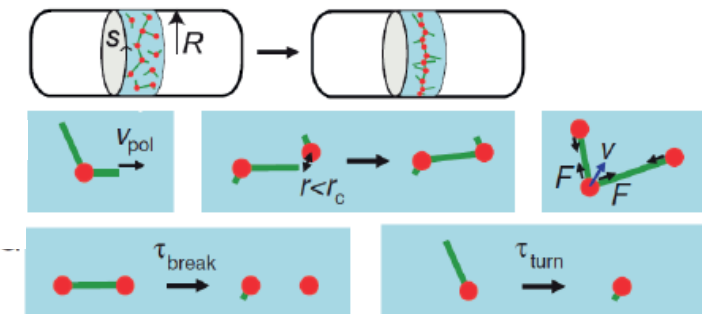

C

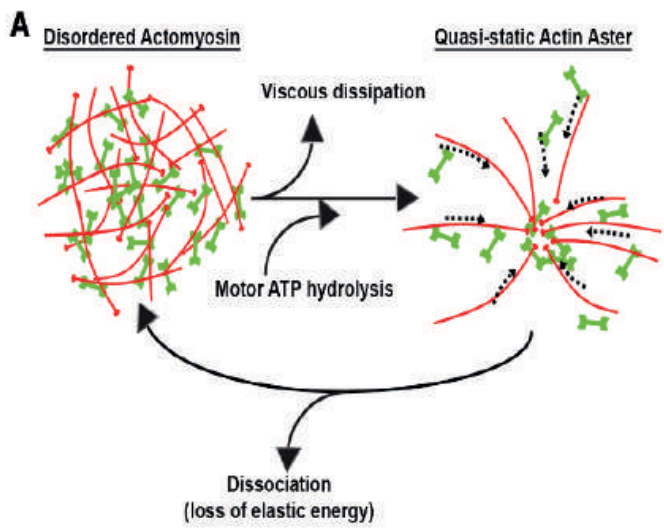

B Continuous description based on component interactions

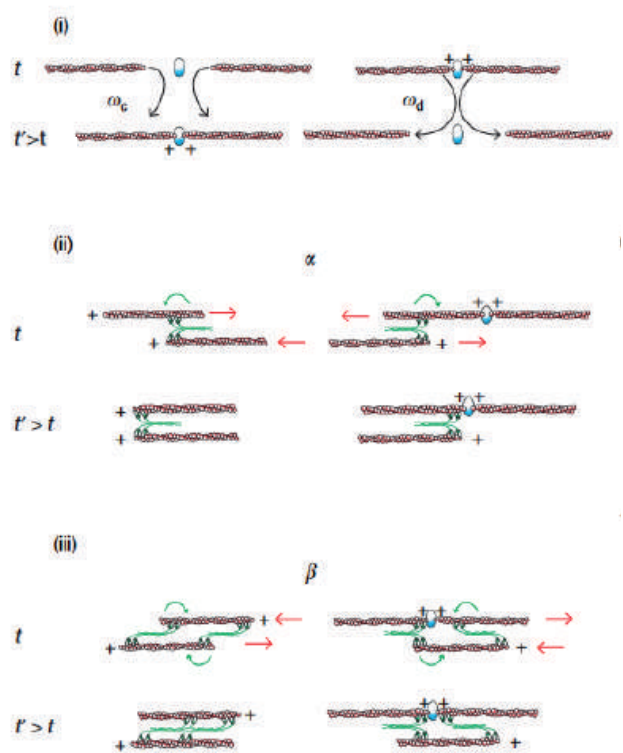

Figure 7

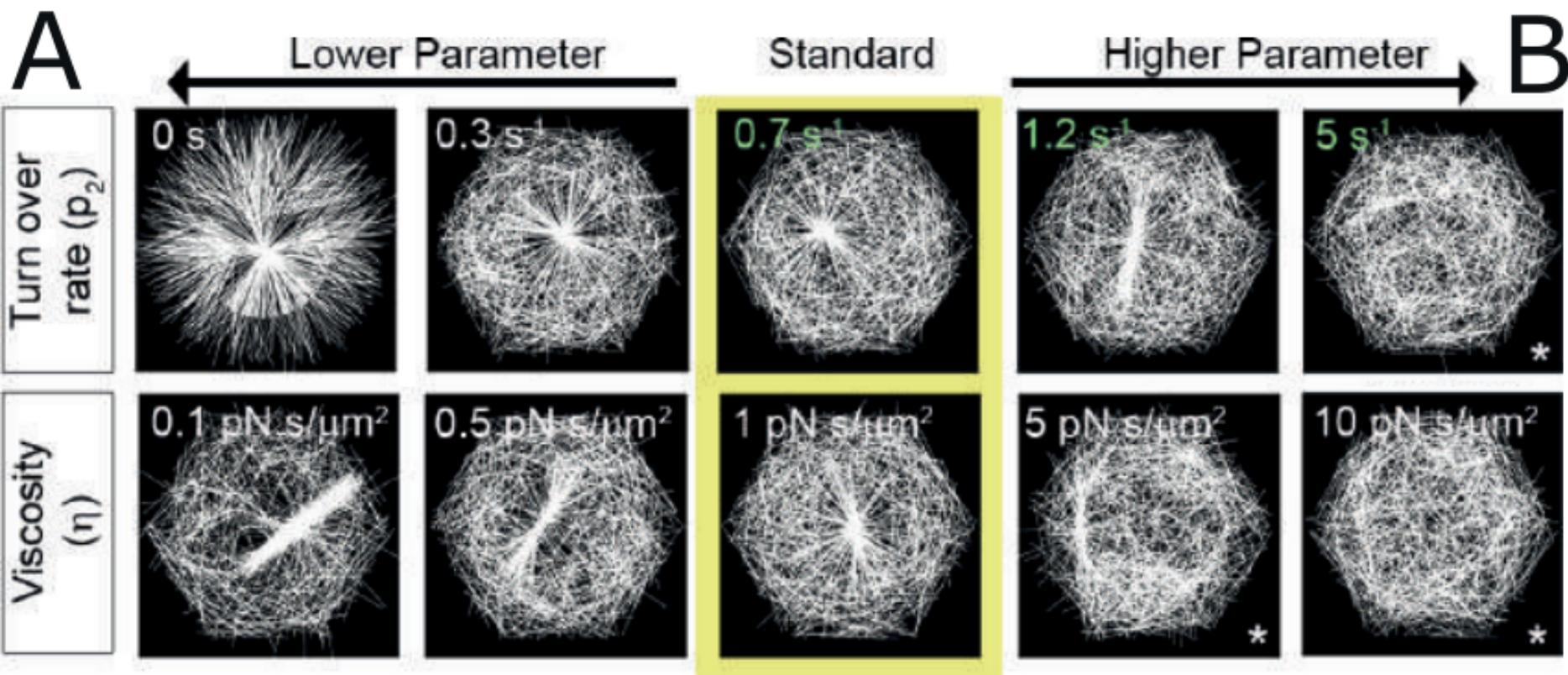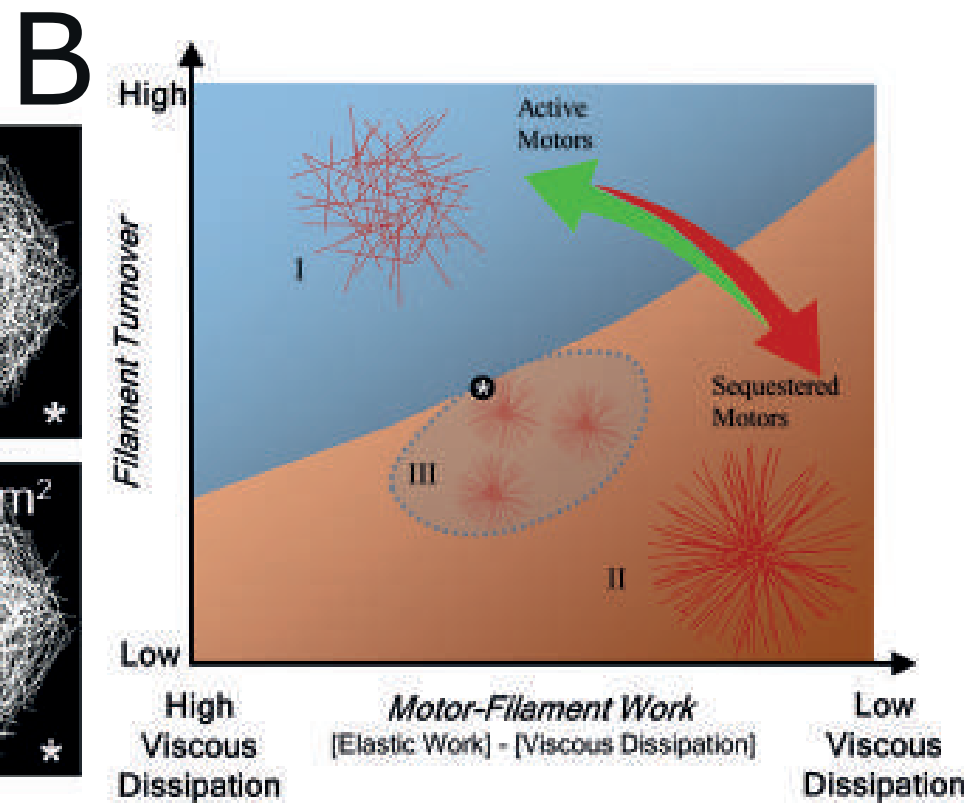

Figure 8